\documentclass[conference]{IEEEtran}
\IEEEoverridecommandlockouts
\usepackage{cite}
\usepackage{amsmath,amssymb,amsfonts}
\usepackage{algorithmic}
\usepackage{graphicx}
\usepackage{textcomp}
\usepackage{xcolor}

\usepackage{booktabs} 
\usepackage{caption}
\usepackage{subfigure}
\usepackage{url}
\usepackage{comment}
\makeatletter \def\mdseries@tt{m} \makeatother
\usepackage{multicol}
\usepackage{multirow}
\usepackage{threeparttable}
\usepackage{enumitem}
\usepackage[cmtip,all]{xy}
\usepackage{listings}

\usepackage[ruled]{algorithm2e}

\definecolor{pink}{cmyk}{0, 0.7808, 0.4429, 0.1412}

\newcommand{\codename}[0]{\textsc{Cloak}\xspace}
\newcommand{\myparagraph}[1]{
	\noindent{\textbf{#1.}}
}

\newcommand{\eg}{\hbox{\emph{e.g.}}\xspace}

\newcommand{\ie}{\hbox{\emph{i.e.}}\xspace}

\newcommand{\etc}{\hbox{\emph{etc.}}\xspace}

\newcommand{\longsquiggly}{\xymatrix{{}\ar@{~>}[r]&{}}}

\newcommand{\owner}[1]{{\bfseries \textcolor{pink}{\small #1}}}

\newcommand{\code}[1]{{\texttt{\small #1}}}

\definecolor{verylightgray}{rgb}{.97,.97,.97}

\lstdefinelanguage{Solidity}{
	keywords=[1]{anonymous, assembly, assert, balance, break, call, callcode, case, catch, class, constant, continue, constructor, contract, debugger, default, delegatecall, delete, do, else, emit, event, experimental, export, external, false, finally, for, function, gas, if, implements, import, in, indexed, instanceof, interface, internal, is, length, library, log0, log1, log2, log3, log4, memory, modifier, new, payable, pragma, private, protected, public, pure, push, require, return, returns, revert, selfdestruct, send, solidity, storage, struct, suicide, super, switch, then, this, throw, transfer, true, try, typeof, using, value, view, while, with, addmod, ecrecover, keccak256, mulmod, ripemd160, sha256, sha3}, 
	keywordstyle=[1]\bfseries,
	keywords=[2]{address, bool, byte, bytes, bytes1, bytes2, bytes3, bytes4, bytes5, bytes6, bytes7, bytes8, bytes9, bytes10, bytes11, bytes12, bytes13, bytes14, bytes15, bytes16, bytes17, bytes18, bytes19, bytes20, bytes21, bytes22, bytes23, bytes24, bytes25, bytes26, bytes27, bytes28, bytes29, bytes30, bytes31, bytes32, enum, int, int8, int16, int24, int32, int40, int48, int56, int64, int72, int80, int88, int96, int104, int112, int120, int128, int136, int144, int152, int160, int168, int176, int184, int192, int200, int208, int216, int224, int232, int240, int248, int256, mapping, string, uint, uint8, uint16, uint24, uint32, uint40, uint48, uint56, uint64, uint72, uint80, uint88, uint96, uint104, uint112, uint120, uint128, uint136, uint144, uint152, uint160, uint168, uint176, uint184, uint192, uint200, uint208, uint216, uint224, uint232, uint240, uint248, uint256, var, void, ether, finney, szabo, wei, days, hours, minutes, seconds, weeks, years},	
	keywordstyle=[2]\color{teal}\bfseries,
	keywords=[3]{block, blockhash, coinbase, difficulty, gaslimit, number, timestamp, msg, data, gas, sender, sig, value, now, tx, gasprice, origin},	
	keywordstyle=[3]\color{violet}\bfseries,
	keywords=[4]{k, @k, p, @p, x, @x, @me, @tee, @all, @winner}, 
	keywordstyle=[4]\color{pink}\bfseries,
	keywords=[5]{reveal},	
	keywordstyle=[5]\color{pink}\bfseries,
	identifierstyle=\color{black},
	sensitive=false,
	comment=[l]{//},
	morecomment=[s]{/*}{*/},
	commentstyle=\color{gray}\ttfamily,
	stringstyle=\color{red}\ttfamily,
	morestring=[b]',
	morestring=[b]"
}

\colorlet{punct}{red!60!black}
\definecolor{background}{HTML}{EEEEEE}
\definecolor{delim}{RGB}{20,105,176}
\colorlet{numb}{magenta!60!black}

\lstdefinelanguage{json}{
    basicstyle=\footnotesize\ttfamily,
    numbers=left,
    numberstyle=\footnotesize,
    stepnumber=1,
    numbersep=9pt,
    showstringspaces=false,
    captionpos=b,
	mathescape=true,
	tabsize=2,
	showtabs=false,
    literate=
     *{0}{{{\color{numb}0}}}{1}
      {1}{{{\color{numb}1}}}{1}
      {2}{{{\color{numb}2}}}{1}
      {3}{{{\color{numb}3}}}{1}
      {4}{{{\color{numb}4}}}{1}
      {5}{{{\color{numb}5}}}{1}
      {6}{{{\color{numb}6}}}{1}
      {7}{{{\color{numb}7}}}{1}
      {8}{{{\color{numb}8}}}{1}
      {9}{{{\color{numb}9}}}{1}
      {:}{{{\color{punct}{:}}}}{1}
      {,}{{{\color{punct}{,}}}}{1}
      {\{}{{{\color{delim}{\{}}}}{1}
      {\}}{{{\color{delim}{\}}}}}{1}
      {[}{{{\color{delim}{[}}}}{1}
      {]}{{{\color{delim}{]}}}}{1},
}

\def\BibTeX{{\rm B\kern-.05em{\sc i\kern-.025em b}\kern-.08em
    T\kern-.1667em\lower.7ex\hbox{E}\kern-.125emX}}
    
\begin{document}

\title{Demo: \textsc{Cloak}: A Framework For Development of Confidential Blockchain Smart Contracts
}

\author{
\IEEEauthorblockN{Qian Ren$^*\quad$ Han Liu$^*\quad$ Yue Li$^*\quad$ Hong Lei$^{*\dagger}$}
\IEEEauthorblockA{
\textit{$^*$Oxford-Hainan Blockchain Research Institute}\\
Hainan, China}
\IEEEauthorblockA{
\textit{$^\dagger$School of Computer Science and Cyberspace Security, Hainan University}\\
Hainan, China}
}

\maketitle

\lstset{numbers=left, numberstyle=\tiny, stepnumber=1, numbersep=4pt}

\begin{abstract}
    In recent years, as blockchain adoption has been expanding across a wide range of domains, \eg, digital asset, supply chain finance, \etc, the confidentiality of smart contracts is now a fundamental demand for practical applications. However, while new privacy protection techniques keep coming out, how existing ones can best fit development settings is little studied. Suffering from limited architectural support in terms of programming interfaces, state-of-the-art solutions can hardly reach general developers.

In this paper, we proposed the \codename framework for developing confidential smart contracts. The key capability of \codename is allowing developers to implement and deploy practical solutions to \emph{multi-party transaction} (MPT) problems, \ie, transact with secret inputs and states owned by different parties by simply \emph{specifying} it.
To this end, \codename introduced a domain-specific annotation language for declaring privacy specifications and further automatically generating confidential smart contracts to be deployed with trusted execution environment (TEE) on blockchain. 
In our evaluation on both simple and real-world applications, developers managed to deploy business services on blockchain in a concise manner by only developing \codename smart contracts whose size is less than 30\% of the deployed ones.

\end{abstract}


\section{Introduction} \label{sec:introduction}
    
With the rapid development of both permissionless and permissioned blockchains, privacy issues have now become 
one of the top concerns for smart contracts, \ie, keep transaction input and contract states as secrets to 
non-relevant participants. In many of the practical applications, privacy is an essential property to achieve, 
\eg, avoid malicious arbitrage on cryptocurrency, protect sensitive information in a cooperative business \etc. 
Unfortunately, despite the importance of smart contract privacy, most of the existing blockchains are 
designed \emph{without privacy} by nature~\cite{wood2014ethereum}. For example, miners of Ethereum verify transactions in a block 
by re-executing them with the exact input and states. Consequently, private data is shared in the entire network.

\begin{figure}[h]
    \centering
    \includegraphics[width=7cm]{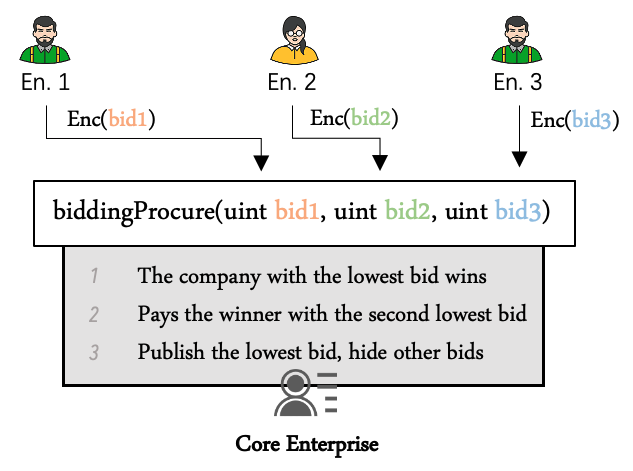}
    \caption{A simplified multi-party transaction scenario}
    \label{fig:vmpt-definition}
\end{figure}

\begin{figure*}[!t]
  \centering
  \includegraphics[width=17.5cm]{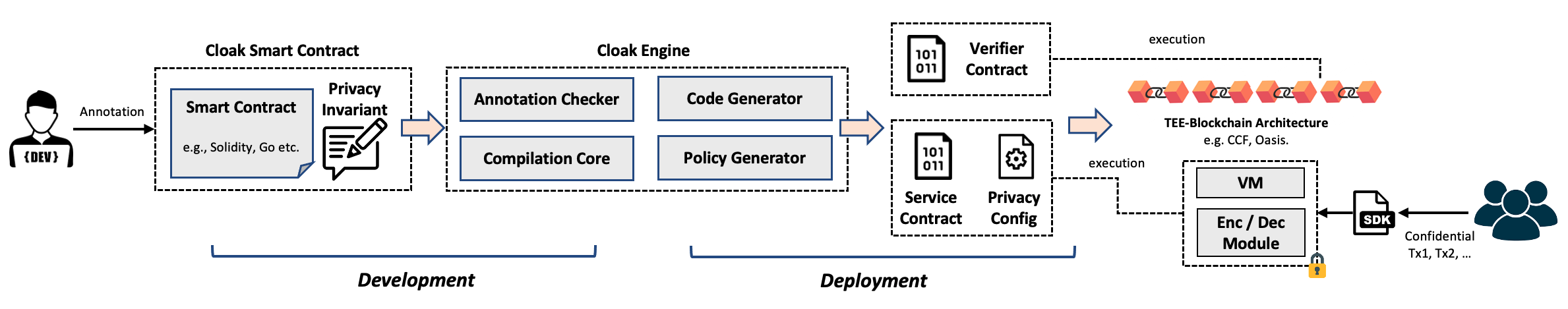}
  \caption{The overall workflow of \codename}
  \label{fig:framework}
\end{figure*}

\myparagraph{\emph{Confidential Smart Contract}} To address the aforementioned problem, researchers have proposed 
a variety of solutions in recent years to the design of \emph{confidential smart contract}. In general, 
these approaches fall into two categories based on cryptography techniques and trusted hardware, respectively. 
For the former class of approaches, techniques including ring signature, homomorphic encryption and 
zero-knowledge proof (ZKP) are adopted to achieve anonymity and privacy~\cite{zkay:CCS2019, Hawk:SP2016, Zether:2019, ZEXE:SP20}. For the latter, Trusted Execution Environment (TEE), \eg, Intel SGX, is commonly 
used to provide confidentiality and trustworthiness~\cite{Ekiden:2019, CONFIDE:SIGMOD20}. More specifically, 
TEE is able to reveal sealed transactions and execute them in enclaves to hide input and contract states with a 
verifiable endorsement from the hardware manufacture.

\myparagraph{\emph{Limitations}}
However, while both classes of solutions provide architectural capabilities to enforce confidential lifecycles 
of transactions, they are non-sufficient for the \emph{development} of practical 
applications. Figure~\ref{fig:vmpt-definition} describes a scenario of procurement bidding among multiple 
enterprises in the setting of supply chain applications. Specifically, each participant submits its 
secret bid and a core enterprise decides a winner with the lowest bid. The core enterprise 
pays the second-lowest bid to the winner instead of the lowest one through updates on its balance. 
For cryptography-based solutions, developers are required to implement a set of off-chain multi-party 
computation programs and on-chain verification smart contracts, as indicated by~\cite{zkay:CCS2019}. On the other hand, 
TEE-based solutions allow developers to implement general smart contracts with secrets owned by only one side 
in a single transaction. Consequently, the implementation needs to process one source of confidential 
bid input at a time, cache intermediate bids and generate final states when bidding completes. To sum up, 
confidential smart contracts in the literature can hardly fit in the practical development of 
multi-party applications.

\myparagraph{\emph{Multi-party Transaction}} In this demo paper, we formalized the Multi-party 
Transaction (MPT) problem on blockchain for the first time and designed the \codename framework as a 
practical solution to it. 
\codename enables developers to develop and deploy confidential smart contracts in an MPT application by 
simply specifying it. Specifically, \codename allows developers to annotate privacy invariants as annotations in contract source code. It checks the privacy specification consistency and then generates deployable smart contract on blockchain. 
The main contributions of this work are as follows: 
\begin{itemize}[leftmargin=*, parsep=1mm, topsep=1mm, partopsep=0mm]
    \item We formalized the \emph{Multi-party Transaction} (MPT) problem on blockchain for the first time, which transacts with secret inputs  and states owned by different parties.
 
    \item We developed a framework, \codename, which allows developers to annotate privacy invariants, including MPT, as annotations in smart contracts, and generate privacy-compliant deployable code.
     
    \item We conducted a preliminary evaluation on real-world contracts with different privacy scenarios. The result shows the easy to use, high efficiency, and low cost of \codename.
\end{itemize}

\myparagraph{\emph{Demonstration Plan}} Our demonstration will showcase the capabilities of \codename in handling real-world data privacy issues. The detailed plan includes i) an automatic type checking, compilation, and deployment process, ii) in-depth explanation of the domain-specific annotation language, generated deployable code, and debug skills when developing with \codename iii) more comparative tests on representative contracts and privacy scenarios.

\section{Multi-party Transaction} \label{sec:vmpt}
    We propose a new smart contract data privacy problem in blockchain called Multi-party Transaction (MPT): For $n(n\in \mathbf{Z^*} \wedge n>1)$ parties, an MPT takes input $x_i$ from each party $i$ and $C(s)$, which is the cryptography commitment of contract old state $s$, \eg, the hash or encryption result of $s$, \etc. Then, it runs the specified function $f$, publishes committed output $r_i$, a proof $p$ and committed contract new state $C(s')$.
$$f(x_1, x_2, ..., x_n, C(s)) \Rightarrow C(r_1), C(r_2), ..., C(r_n), C(s'), p$$
An MPT should satisfy following two attributes:

\begin{itemize}[leftmargin=*]
    \item \textit{Confidentiality}: Each party $i$ knows $r_i$ without knowing $\{x_j, r_j | i\neq j\}$ except what can be derived from $x_i, r_i$ itself. $i$ should also know state $s$ or $s'$ only when it's owned and provided by him.
    \item \textit{Verifiability}: With $p$, all nodes could verify that the commitment of new state $C(s')$ and return value $C(r_i)$ is the correct result of a function $f$, which takes unknown $\{x_j|j=1..n \wedge j \neq i\}$ from $n(n\in \mathbf{Z^*} \wedge n>1)$ parties and old state $s$, which is committed by on-chain $C(s)$.
\end{itemize}

MPT is different from Multi-party Computation (MPC). In MPC, even though all MPC participants acknowledge the transaction and record the result on a blockchain, it is hard for other nodes to verify it. Consequently, other nodes regard it as normal immutable data, making the MPC results lose widespread trust. In contrast, MPT achieves the same level of security and final consistency as smart contracts with proofs.


\section{The \codename Framework} \label{sec:cloak}
    To handle the MPT problem, we designed a \codename framework. Figure~\ref{fig:framework} shows the workflow of \codename. It mainly divides into two phase, development and deployment. In the development phase, developers first annotate privacy invariants in Solidity smart contract intuitively to get \codename smart contract. \code{Annotation Checker} checks the annotation to make sure the privacy invariants are correct. The core of the development phase is \code{Cloak Engine}, in which the \code{Code Generator}, \code{Policy Generator}, and \code{Compilation Core} generate verifier contract, service contract, and privacy config. All generated code will be deployed to blockchains with TEE-Blockchain Architecture, \eg, Oasis~\cite{Ekiden:2019}, CCF~\cite{ccf2019}, \etc.

\subsection{Develop Confidential Smart Contract}
\myparagraph{Annotate Privacy Invariants}
Developers could annotate variable owner in the declaration statement to one of the \{\owner{all}, \owner{me}, \owner{id}, \owner{tee}\}. The \owner{all} means public; \owner{me} means the \code{msg.sender}; \owner{id} is declared variable in type \code{address}; \owner{tee} means any registered address of SGX with \codename runtime. 

With \codename, users could intuitively specify the MPT in Figure~\ref{fig:vmpt-definition} as a \codename smart contract, the \textit{.cloak} file in Listing~\ref{listing:cloak}. In line 1, the developer could declare the key of \code{balances} as a temporary variable \owner{k}, then specifies the corresponding value is owned by the account with address \code{k}, \eg, \code{balances[tenderer]} is only known by the \code{tenderer} in line 23. In line 2, the developer specifies \code{mPrice} should be public. In line 6-7, to handle an uncertain number of suppliers, the developer declares owners \owner{p} and specifies the owners' owned data separately in two dynamic arrays. In line 10, the return value \code{sPrice} is owned by the \code{winner}. In line 12-13, the developer \owner{reveal} private data to another owner, which is forced by \codename to avoid unconsciously leaking privacy. In line 14-24, it computes the lowest price, the second lowest price, and the winner. The computation is based on the operation between private data from different parties, \eg, \code{bids[i] < sPrice}, \code{ balances[tenderer] += sPrice}.

\begin{lstlisting}[language=Solidity, caption=\codename smart contract of bidding procurement, label=listing:cloak]
contract SupplyChain {
    mapping(address !k => uint @k) balances;
    uint @all mPrice;
    
    function biddingProcure(
        address[!p] parties, 
        uint[@p] bids,
        address tenderer
    ) public 
    returns (address winner, uint @winner sPrice) {
        winner = parties[0];
        uint mPrice = reveal(bids[0], all);
        sPrice = reveal(bids[0], winner);
        for (uint i = 1; i < parties.length; i++) {
            if (bids[i] < mPrice) {
                winner = parties[i];
                sPrice = mPrice;
                mPrice = bids[i];
            } else if (bids[i] < sPrice) {
                sPrice = bids[i];
            }
        }
        balances[tenderer] -= sPrice;
        balances[winner] += sPrice;
    }
}

\end{lstlisting}

\myparagraph{\emph{Annotation Checker}}
Taking a \codename smart contract, \codename ignores the annotation to checks the Solidity validation first. Then, \codename builds an Abstract Syntax Tree (AST) for further analysis. It infers data owner and checks the privacy invariants. It traversals the AST in post-order and updates each parent node's owner $o_p=o_l\cup o_r$. The $o_l$ and $o_r$ is the owner set of the left and right child node respectively. \codename recognizes a function as an MPT if $TEE\in o$ or $|o\setminus\{all\}|\geq2$. The latter means the function will take private data from different parties. Then, \codename checks privacy invariants consistency. For example, \codename prohibits developers from implicitly assigning their private data to variables owned by others.

\subsection{Deploy Confidential Smart Contract}
\myparagraph{\emph{Policy Generator}}
With checked AST, \code{Policy Generator} generates a privacy config $P$ for the contract. $P$ simplifies and characterizes the privacy invariants. Typically, $P$ includes variables with data type and owners. It also includes ABI, a read-write set of each function. Specifically, $P$ records each function's characteristics from four aspects, \textit{inputs}, \textit{read}, \textit{mutate} and \textit{return}. The \textit{inputs} includes its parameters with specified data type and owner; \textit{read} records state variables the function read in execution; \textit{mutate} records the contract states it mutated; \textit{return} records the return variables. Since $P$ has recorded the details of state variables in the head, \eg, data type and owner, \code{Policy Generator} leaves the variable identities in \textit{read}, \textit{mutate} and \textit{return}. 

\myparagraph{\emph{Code Generator}}
\code{Code Generator} generates a service contract $F$ and a verifier contract $V$. While leaving the computation logic in $F$, \code{Code Generator} generates $V$ to verify the result and update the state. In $V$, \code{Code Generator} first imports a pre-deployed \codename TEE registration contract, which holds a list of registered SGXs with \codename runtime. Then \codename transforms each MPT function in \textit{.cloak} into a new function in $V$, which verifies the MPT proof $p$ and assigns new state $C(s')$ later. 

\myparagraph{\codename Client}

With configured nodes IP and ports by developers, \codename deploys the confidential smart contract \textbf{runtime} to TEE-Blockchain Architecture to get trusted \codename executors $E$s. The \textbf{runtime} includes \code{VM} and a \code{Enc/Dec Module}. Then, \codename deploys a SGX registration contract on the blockchain and registers the $E$s' certificate. For each \codename smart contract, \codename will deploys generated $P$ and $F$ to $E$s and $V$ to the blockchain separately.

When a participant proposes an MPT, each participant $i$ provides the $x_i$ to SDK. $x_i$ will be encrypted and sent to $E$s. According to deployed $P$, $E$s wait for all private \code{inputs}, synchronize the \code{read} state, construct a transaction, and execute it in enclaves. Then, $E$s encrypt return values and mutated states according to \code{return} and \code{mutate}. Finally, $E$s announce a result transaction on-chain with an MPT proof $p$. The $p$ is $E$s' signature, \ie, $p=Sig_E<P, F, C(s), C(r_i), C(s')>$. It  means compliant to $P$, $E$s confidentially execute $F$ with private inputs $x_i$ and old state $s$ committed by $C(s)$ in enclaves, commit return value $r_i$ and new state $s'$ to get result $C(r_i), C(s')$. Upon receiving the announcement transaction with proof, all nodes and $V$ could believe an MPT real happened and get the result.

\section{Preliminary Evaluation} \label{sec:evaluation}
    We conducted a preliminary evaluation of \codename on 9 contracts. The 9 contracts vary from scenarios, privacy invariants and have representative LOC in Ethereum smart contract~\cite{tech2019hegedus}.
\begin{table}[!htbp]
  \centering
  \caption{The LOC of code before and after using \codename. the number of \code{Functions} (public, private, MPT) ; the LOC of \codename smart contract; the whole generated verifier contract $\textbf{V}_{all}$; generated service contract $\textbf{F}$, verifier contract $\textbf{V}$ and privacy config $\textbf{P}$ for MPTs}
  \label{tab:contracts-evaluation}
  \setlength{\tabcolsep}{1.8mm}{
      \begin{tabular}{lcccccc}
        \toprule
        \multirow{2}{*}{\textbf{Name}} & \multirow{2}{*}{\textbf{\#Functions}} & \multirow{2}{*}{\textbf{\#\codename}} & \multirow{2}{*}{\#$\textbf{V}_{all}$} &  \multicolumn{3}{c}{\textbf{MPT related}} \\
        \cmidrule{5-7} & & & & \#$\textbf{F}$ & \#$\textbf{V}$ & \#$\textbf{P}$ \\
        \midrule
        \texttt{PowerGrid} & 4(1, 1, \textbf{2}) & 25 & 146  & 23 & 126 & 72 \\
        \texttt{Bidding} & 4(0, 2, \textbf{2})  & 44 & 148  & 38 & 123 & 102 \\
        \texttt{SupplyChain} & 6(0, 5, \textbf{1})  & 68 & 249 & 36 & 145 & 85 \\ 
        \texttt{Scores} & 6(0, 2, \textbf{4})  & 77 & 239  & 57 & 211 & 174 \\ 
        \texttt{Insurance} & 8(2, 3, \textbf{3}) & 89 & 356  & 52 & 271 & 199  \\
        \texttt{ERC20Token} & 11(4, 4, \textbf{3})  &112 & 347 & 56 & 218 & 173 \\
        \texttt{YunDou} & 14(10, 0, \textbf{4}) & 279 & 501 & 166 & 361 & 345 \\
        \texttt{Oracle} & 22(19, 0, \textbf{3})  & 326 & 413 & 93 & 190 & 196 \\
        \texttt{HTLC} & 39(31, 0, \textbf{8}) & 1029 & 852 & 429 & 401 & 443 \\
        \bottomrule
      \end{tabular}
  }
\end{table}

\myparagraph{Programming Simplicity}
Table~\ref{tab:contracts-evaluation} shows the LOC of privacy-compliant code before and after using \codename.  Specifically, for MPT functions, while developers simply annotate in Solidity contract,  \codename generates a total of 2.97-9.61X LOC, including 0.73-5.47X verifier contract LOC in Solidity and 1.03-3.13X privacy config in JSON. Therefore, \codename significantly reduces the development complexity of privacy in cryptography understanding and code implementation. 

\myparagraph{Performance}
It costs \codename less than 1s to compile 8 contracts, while the biggest contract \texttt{HTLC} takes 5s. This is completely acceptable. 

\begin{figure}[!htbp]
    \centering
    \includegraphics[width=8cm]{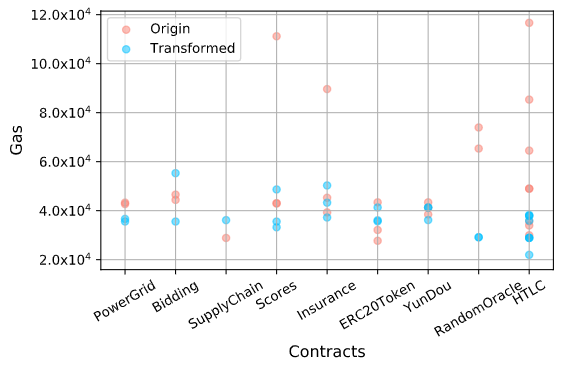}
    \caption{Gas cost of MPT before and after \codename}
    \label{fig:txs-gas-cost}
\end{figure}

\myparagraph{Gas Cost}
In Figure~\ref{fig:txs-gas-cost}, for a total of 27 MPT in 9 contracts, the transformed MPTs cost 0.8X to original no privacy transactions on average. Specifically, the lowest one is 0.35X. The highest cost in the last 8 is 1.25X. Overall, running transactions on transformed contracts are feasible at a moderate cost.

\section{Demonstration Description} \label{sec:demo}
    Currently, for simplicity, we implement the \codename framework for a TEE-Blockchain Architecture consists of Ethereum and SGX simulators. It is a command-line tool and requires dependencies including ANTLR 3.82, Python 3.8, and solc~\cite{solc0517} 0.5.17. A simple command to run \codename is:

\code{\$ cloak -i example.cloak -o out\_dir}

Specifically, \code{cloak} is a running script to start \codename. \code{-i} specify the input \codename smart contract, \textit{.cloak}. \code{-o} specify the output directory to store generated code. In the demonstration, we use two use cases. \ie, basic development and deployment and annotation debugging. The first helps developers understand the basic capabilities of \codename; the second shows fine-grained info of \textit{.cloak} which is critical in debugging.

\myparagraph{\emph{Use Case 1: Basic Development and Deployment}}
Developers use \codename on a supplied \codename smart contract, \ie, a contract with annotated private data and data owner. First, we go through the source code of the specified contract with developers and introduce its metadata. Next, a participant will start the compilation of \codename with the basic configuration. We let the participant monitor the runtime logs generated by \codename in annotation checking. When it finishes policy and code generation, \codename will display an overview of the compilation process, \eg, the private data and its owner, the function type recognized, the time used, \etc. Furthermore, we go into details of each function with developers and explain why \codename recognizes the function as a public, private, or MPT function. The developers can see the specific privacy statement resulting in the recognized function type and check whether the \code{Privacy Config}  is expected.

Additionally, by using \code{SDK}, developers could deploy the generated verifier contract on a local Ganache-driven blockchain, the runtime, service contract, and privacy config in an SGX simulator. When it's done, they can send an MPT cooperating with two parties to know how it works and make sure it is privacy compliant.

\myparagraph{\emph{Use Case 2: Annotation Debugging}}
This case focuses on debugging privacy invariant annotation when developing \codename smart contracts. We provide a group of \codename smart contracts with different errors. Developers can learn about debugging by flexibly using different commands to fine-grained control the \codename behavior. For example, developers could additionally use \code{-s/--solc} to ignore the \codename annotation and check the Solidity validation. \code{-t} specify the \codename just doing annotation checking and report privacy config. \code{--debug} specify \codename to show more compilation details, \ie, the time used in annotation checking and code generation of each function, the hash of generated verifier contract, service contract, privacy config, and runtime.

\bibliographystyle{plain}
\bibliography{bibliography}

\end{document}